\documentclass[aps,prl,nofootinbib,twocolumn]{revtex4}
\usepackage{epsfig}
\usepackage{slashed}
\usepackage{xcolor}

\newcommand{\bel}[1]{\begin{equation}\label{#1}}
\newcommand{\bal}[1]{\begin{eqnarray}\label{#1}}

\newcommand{\be}{\begin{equation}}
\newcommand{\ee}{\end{equation}}
\newcommand{\ba}{\begin{eqnarray}}
\newcommand{\ea}{\end{eqnarray}}

\newcommand{\bes}{\begin{equation*}}
\newcommand{\ees}{\end{equation*}}

\begin{document}
\title{Large $N_c$ Deconfinement Transition in the Presence of a Magnetic Field}

\author{Eduardo S. Fraga$^1$, Jorge Noronha$^2$ and Let\'\i cia F. Palhares$^3$}

\affiliation{$^1$Instituto de F\'\i sica, Universidade Federal do Rio de Janeiro, 
Caixa Postal 68528, Rio de Janeiro, RJ 21941-972, Brazil \\ 
$^2$Instituto de F\'{i}sica, Universidade de S\~{a}o Paulo, C.P. 66318, 05315-970 S\~{a}o Paulo, SP, Brazil \\ 
$^3$Instituto de F\'\i sica, Universidade do Estado do Rio de Janeiro, 
Rua S\~ao Francisco Xavier 524,  Maracan\~a, Rio de Janeiro, RJ 20550-013, Brazil}

\begin{abstract}
We investigate the effect of a homogeneous magnetic field on the thermal deconfinement transition of QCD in 
the large $N_c$ limit. 
First we discuss how the critical temperature decreases due to the inclusion of $N_f \ll N_c$ flavors of massless quarks in comparison to the pure glue case. Then we study the equivalent correction in the presence of an external Abelian magnetic field. 
To leading order in $N_{f}/N_{c}$, the deconfinement critical temperature decreases with the 
magnetic field if the flavor contribution to the pressure behaves paramagnetically, with a sufficiently 
large magnetization as to overcome any possible magnetic effects in the 
string tension. Finally, we discuss the effects 
from a finite quark mass and its competition with magnetic effects.
\end{abstract}

\maketitle


{\it Introduction.} The phase diagram of strong interactions in the presence of a classical, constant, 
and uniform magnetic background has been attracting increasing interest in the last few years. Strong 
(Abelian) magnetic fields not only provide another control parameter to probe the phase structure of QCD   
but are also currently generated in non-central ultrarelativistic heavy ion collisions at RHIC-BNL and at the LHC-CERN. In fact, these fields are believed to be the largest ever produced since the times of the electroweak phase transition in the early universe, reaching values on the order of $B \sim 10^{19}~$Gauss ($eB \sim6\,m_{\pi}^2$) and even much higher \cite{magnetic-HIC}. Furthermore, lattice Monte Carlo simulations are not constrained 
by the sign problem in this case and can produce a trustworthy $T \times eB$ phase diagram, among other 
results. Nevertheless, the mapping of this new phase diagram is still in its infancy and presents some 
conflicting pictures coming from different model calculations. 

In this paper we study the behavior of the deconfining critical temperature $T_c$ in the presence of a strong 
magnetic field in the large $N_{c}$ limit of QCD. This provides a well-defined setup for a clean, 
semi-quantitative description by essentially counting powers of $N_{f}/N_{c}$ (with $N_f$ being the number of quark flavors) when matching pressures 
for the confined and deconfined sectors. Our analysis suggests that the deconfinement 
temperature decreases with the magnetic field for small $N_{f}/N_{c}$, provided that the flavor contribution to the large $N_c$ pressure is paramagnetic. We also discuss how the critical temperature for the pure glue theory decreases due to the leading order correction in $N_{f}/N_{c}$ in the absence of a magnetic field. 

All model calculations so far have suggested that sufficiently large magnetic fields, typically 
$eB \sim10\,m_{\pi}^2$, could bring remarkable modifications in the QCD phase diagram, from shifting 
the chiral and the deconfinement phase transition 
lines \cite{Agasian:2008tb,Fraga:2008qn,Menezes:2008qt,Boomsma:2009yk,Fukushima:2010fe,Avancini:2011zz,Kashiwa:2011js,Chatterjee:2011ry,Andersen:2011ip,Skokov:2011ib,Fukushima:2012xw} 
to transforming the vacuum into a superconducting medium via $\rho$-meson 
condensation \cite{Chernodub:2010qx}. In particular, most model descriptions have predicted either 
an increase or a flat behavior for the deconfinement critical line as $eB$ is increased to very large values. 
Exceptions can be found in Ref. \cite{Agasian:2008tb}, where the critical temperature vanishes at a finite 
critical value of $eB_{c}\sim 25m_{\pi}^2$, featuring the disappearance of the confined phase at large 
magnetic fields, and in \cite{Fraga:2008qn}, where vacuum corrections are disregarded, 
and $T_{c}$ diminishes with $eB$.

The first pioneering lattice simulations \cite{lattice-delia}, still with large values for the pion mass, also 
suggested a very mild increase of the critical temperature with $eB$. However, recent lattice simulations 
with physical masses \cite{Bali:2011qj} have shown that the critical temperature for deconfinement 
actually falls as the magnetic field increases. However, instead of falling with a rate that will bring it to zero 
at a given critical value of $eB$, it falls less and less rapidly, tending to saturate at large values of $B$ 
in agreement with what one would expect from the phenomenon of magnetic catalysis \cite{Gusynin:1994xp,Bali:2012zg}. 
An exercise within the MIT bag model with the appropriate treatment of the subtleties of renormalization 
at finite $B$ has shown remarkable qualitative agreement with these lattice findings with respect to 
the behavior of $T_{c}(eB)$, i.e. it decreases and saturates for very large fields \cite{Fraga:2012fs}. To 
the best of our knowledge, even if known to be crude in numerical precision and missing the correct nature 
of the (crossover) transition, this is the only description to date that captures the correct qualitative behavior of 
the deconfining transition in a magnetic background.

Although a description of the deconfinement transition in the presence of an external magnetic field in 
terms of the MIT bag model is, of course, very simple, we believe it encodes an essential 
ingredient to provide a qualitative description of the behavior of $T_{c} \times eB$: confinement. The fact that the MIT bag model incorporates 
confinement (even if in its simplest fashion) seems to make it suited to describe the behavior of $T_{c}$ 
as a function of external parameters, as hinted by a previous successful description of the behavior of 
the critical temperature as a function of the pion mass and isospin chemical potential, as compared to 
lattice data, where chiral models failed even qualitatively \cite{Fraga:2008be,leticia-tese}. This suggests 
that confinement dynamics may play a central role in guiding the functional behavior of $T_{c}$ and 
points towards a large $N_{c}$ description of the associated magnetic thermodynamics.

{\it Large $N_{c}$ thermodynamics.} The large $N_c$ limit provides a great opportunity to study several aspects of QCD \cite{'tHooft:1973jz,'tHooft:1974hx,Witten:1979kh,Manohar:1998xv}. 
Feynman diagrams are reorganized according to their dependence on $N_c$ and, when $N_c\to \infty$, only planar diagrams are relevant. The theory is still asymptotically free with a perturbative beta function defined in terms of the 't Hooft coupling $\lambda \equiv g^2 N_c$ and a renormalization group invariant energy scale $\Lambda_{QCD}$ at which the associated coupling becomes strong. While confinement has not been proven in this limit, it is widely believed that in the vacuum the physical degrees of freedom are weakly interacting (since interactions go as $1/N_c$), colorless glueballs. $N_f$ quark degrees of freedom in the fundamental representation can be added to this theory and the corresponding mesons are free when $N_c\to \infty$ while baryons become extremely heavy, $M_{baryon}\sim N_c \,\Lambda_{QCD}$
 \cite{Witten:1979kh,Manohar:1998xv}.

Lattice QCD calculations \cite{Teper:2008yi} show that the deconfinement phase transition of pure glue $SU(N_c)$ gauge theory becomes first order
 when $N_c \geq 3$ \cite{Boyd:1996bx,muitaslattice,Lucini:2005vg,Borsanyi:2012ve} with a critical temperature 
$\lim_{N_c\to\infty}T_c/\sqrt{\sigma_0}=0.5949(17) + 0.458(18)/N_c^2$ \cite{Lucini:2012wq}, where $\sigma_0 \sim (440\, {\rm MeV})^2$ is the string tension of the large $N_c$ pure glue theory. The thermodynamic properties of pure glue do not seem to change appreciably when $N_c \geq 3$ \cite{morelattice,Panero:2009tv}, which suggests that 
large $N_c$ arguments may indeed capture the main physical mechanism behind the deconfinement phase transition of QCD (at least when $N_c$ is sufficiently large).

The fact that $\lim_{N_c\to \infty}T_c/\sqrt{\sigma_0}\sim \mathcal{O}(N_c^0)$ and that the deconfining phase transition becomes strong first order can be readily understood using the following argument \cite{Lucini:2005vg}. When $N_c \to \infty$ and $N_f=0$, in the confined phase glueballs are very weakly interacting and, since they are colorless, they only contribute to the pressure at $\mathcal{O}(N_c^0)$. String breaking processes cannot occur when $N_f=0$. Therefore, when $N_c\to \infty$ the only contribution to the pressure of the confined phase comes from the gluon condensate $\sim N_c^2 \Lambda_{QCD}^4$, which we write in terms of the renormalization group invariant $\sigma_0$ as $P_{conf}= c_0^4 N_c^2 \sigma_0^2$, where $c_0$ is a positive number of order 1. 
Moreover, it should be noticed that the entropy density in the confined phase vanishes.

On the other hand, asymptotic freedom implies that in the planar limit the gluon pressure is $P_{gluon}(T)=N_c^2 T^4 \,c_{SB}^4\, f_{glue}(T/\sqrt{\sigma_0})$, where $c_{SB}$ is a positive constant determined from the Stefan-Boltzmann limit and  
$\lim_{T/\sqrt{\sigma_0}\to \infty}f_{glue}(T/\sqrt{\sigma_0})=1$. The function 
$f_{glue}$ depends implicitly on the 't Hooft coupling $\lambda(T)$ and, while its general form is not known when $T \sim \sqrt{\sigma_0}$, thermodynamical equilibrium imposes that it should be a monotonically
increasing function of $T$ that interpolates from 0 when $T\to 0$ to 1 for $T\to \infty$. Its form can be computed using perturbation theory at sufficiently high temperatures 
where $\lambda$ becomes very small \cite{kapustagale}. If $N_f=0$, since the pressure 
is always continuous at any phase transition, we see that there must be a deconfinement critical temperature
defined by the condition $P_{glue}\left(T_c^{(0)}/\sqrt{\sigma_0}\right) = P_{conf}$ or, equivalently,
\begin{equation}
 c_0^4 N_c^2\,\sigma_0^2=N_c^2 T_c^{(0)\,\,4}\,c_{SB}^4\, f_{glue}\left(T_c^{(0)}/\sqrt{\sigma_0}\right)\,,
\end{equation}
which implies that the solution $T_c^{(0)}$ is a pure number of $\mathcal{O}(N_c^0)$ that in general cannot be computed perturbatively since it is obtained from the self-consistent equation
\begin{equation}
\frac{T_c^{(0)}}{\sqrt{\sigma_0}}\,f_{glue}^{1/4}\left(\frac{T_c^{(0)}}{\sqrt{\sigma_0}}\right)=\frac{c_0}{c_{SB}}\,.
\label{firstTc}
\end{equation}
Since $f_{glue}$ increases monotonically with $T$ one obtains that $T_c^{(0)}$ must increase with $c_0$ (note that the critical temperature only vanishes if $c_0 \to 0$) \cite{comment1}. 
Lattice calculations have shown that $T_c^{(0)}/\sqrt{\sigma_0} \sim 0.59$ \cite{Lucini:2012wq}. 
The phase transition to a $Z_{N_c}$ symmetric deconfined phase is then of first order when $N_c \to \infty$, $N_f=0$, and the entropy density 
jumps from zero to a finite number of $\mathcal{O}(N_c^2)$ at $T_c^{(0)}$. 

{\it  Leading $N_f/N_{c}$ corrections.} 
The first correction to this picture appears with the inclusion of $N_f$ flavors of {\it massless} quarks. The previous $Z_{N_c}$ symmetry is broken explicitly in 
the deconfined phase because of the presence of quarks. While the $U(N_f)\otimes U(N_f) \to U(N_f)_{vector}$ pattern of (spontaneous) symmetry breaking 
leads to $N_f^2-1$ Goldstone bosons (the ``pions"), their contribution to the pressure of the confined phase is of $\mathcal{O}(N_f^2 N_c^0)$, being negligible when $N_f\ll N_{c}$. 

The presence of quark flavors, even in the massless limit, can lead to corrections of order $\sim N_f N_c$ to the vacuum pressure. In the double line notation \cite{'tHooft:1973jz}, the addition of quark flavors leads to diagrams with boundaries and it is possible to write down an infinite series of diagrams (each one with a power of $\lambda$) that can enter at that order due to production of quark-antiquark loops. Once quark loops can appear in the theory, it is natural to assume that the value of the string tension decreases with the leading $N_f/N_c$ correction with respect to the $N_f=0$ value. This occurs because $q\bar{q}$ pairs can now be produced, which should decrease the linear confining potential experienced by infinitely massive probes in the fundamental representation (i.e., the heavy quark potential). Therefore, we assume that the string tension in the presence of the leading flavor correction is given by $\sigma/\sigma_0=1-\alpha N_f/(2N_c)$, where $\alpha$ is {\it positive definite}. Given this expression for the string tension, the large $N_c$ vacuum pressure becomes, in the presence of massless quarks,
\begin{equation}
P_{conf}=c_0^4 N_c^2\,\sigma_0^2\left(1-\alpha \frac{N_f}{N_c}\right)\,.
\end{equation} 
When quarks are massive, there is another term of order $N_f N_c$ in the vacuum pressure given by the quark condensate contribution to the trace anomaly. We will discuss the massive quark case later; for now we keep the focus on the massless quark limit. 

Once $N_f$ flavors are included in the theory, the deconfined pressure also receives a contribution of order $N_c N_f$, which we denote here by $P_{quark}(T)$. The most general expression for this quantity has the form $P_{quark}(T) = N_c N_f \,T^4\,c_{q\,SB}^4\, f_{quark}(T/\sqrt{\sigma_0}
)$, where $c_{q\,SB}$ is the corresponding positive dimensionless number 
computed in the Stefan-Boltzmann limit and $f_{quark}$ is a monotonically increasing function of $T$ such that $\lim_{T/\sqrt{\sigma_0}\to \infty}f_{quark}(T)=1$.

When $N_f/N_c \ll 1$  the explicit breaking of $Z_{N_c}$ 
symmetry is small, slightly smoothening the phase transition into a very rapid crossover.  The Polyakov loop below the transition is small, i.e., of order $N_f/N_c$. (This is why the contribution from a Polyakov loop potential to the pressure goes effectively as $\sim N_f^2$, i.e., a meson-like contribution.) The balance equation that defines the critical temperature $T_c^{(1)}$ modified by the quark flavors is 
obtained by equating the pressures $P_{conf}=P_{glue}\left(T_c^{(1)}\right)
+P_{quark}\left(T_c^{(1)}\right)$. Since $f_{quark}$ is a monotonic function of $T$, one should expect that the critical temperature gets shifted towards smaller values. In fact, in the limit where $N_f/N_c \ll 1$ one finds the self-consistent equation
\begin{eqnarray}
\frac{T_c^{(1)}}{\sqrt{\sigma_0}}&=&\frac{c_0}{c_{SB}\,f_{glue}^{1/4}\left(\frac{T_c^{(1)}}{\sqrt{\sigma_0}}\right)}\left[1-\frac{\alpha}{4}\frac{N_f}{N_c}\right.\nonumber \\
&-& \left. \frac{1}{4}\frac{N_f}{N_c}\frac{c_{q\,SB}^4\,f_{quark}\left(\frac{T_c^{(1)}}{\sqrt{\sigma_0}}\right)}
{c_{SB}^4\,f_{glue}\left(\frac{T_c^{(1)}}{\sqrt{\sigma_0}}\right)}\right]\,.
\label{2ndTc}
\end{eqnarray}
It is possible to obtain the effect of the leading order $N_f/N_c$ correction on $T_c^{(1)}$ in terms of $T_c^{(0)}$. 
Keeping only the first correction in $N_f/N_c$, one may take $T_c^{(1)}\mapsto T_c^{(0)}$ inside the brackets in the equation above.
Since the ratio $f_{quark}/f_{glue}$ is positive, one can define a new (still positive) constant given by
\begin{equation}
 c_1(N_f) \equiv c_0 \left[1-\frac{\alpha}{4}\frac{N_f}{N_c}-\frac{1}{4}\frac{N_f}{N_c}\frac{c_{q\,SB}^4\,f_{quark}\left(\frac{T_c^{(0)}}{\sqrt{\sigma_0}}\right)}
{c_{SB}^4\,f_{glue}\left(\frac{T_c^{(0)}}{\sqrt{\sigma_0}}\right)}\right]\,.
\label{c1Nf}
\end{equation}
Therefore, the self-consistent equation for $T_c^{(1)}$ has actually the same form as Eq.\ (\ref{firstTc}) and is
given by 
\begin{equation}
\frac{T_c^{(1)}}{\sqrt{\sigma_0}}\,f_{glue}^{1/4}\left(\frac{T_c^{(1)}}{\sqrt{\sigma_0}}\right)=\frac{c_1(N_f)}{c_{SB}}\,.
\label{newTc}
\end{equation}
Thus, since $c_1(N_f)< c_0$ and $f_{glue}$ is monotonically increasing with $T$, we see that the leading effect of $N_f$ massless flavors in the large $N_c$ limit is to decrease the critical temperature by a small amount of order $N_f/N_c$ with respect to $T_c^{(0)}$. In other words, the addition of a small number light quark flavors should decrease the value of the deconfinement critical temperature at large $N_c$. While the validity of any result obtained in the large $N_c$ limit cannot be straightforwardly extended to the physical $N_c=N_f=3$ case, it is reassuring to know that lattice QCD simulations \cite{Karsch,Aoki:2006we,Aoki:2006br,Bazavov:2011nk} performed with $N_c=3$ have found that light quark flavors decrease the deconfinement temperature. 
      
{\it Large $N_{c}$ behavior of $T_{c}\times (eB)$.} The same line of argument used above can be employed to study what happens to the deconfinement critical temperature in the presence of an external magnetic field in the large $N_c$ limit of QCD. Assuming that $N_f/N_c \ll 1$ and the quark mass $m_q=0$, the magnetic field affects the confined pressure at order $N_f N_c$ via the effects of quark loops (higher order corrections were studied in \cite{Agasian:1999sx,Agasian:2000hw,Cohen:2007bt}). Thus, we promote $\alpha$ to be a function of the magnetic field as follows: $\alpha \to \tilde{\alpha}(eB/\sigma_0)$. While we cannot say anything about the explicit magnetic field dependence of $\tilde{\alpha}$, since it depends on the non-perturbative QCD dynamics, we assume that $\tilde{\alpha}(eB/\sigma_0)$ is still positive definite. The confined pressure to leading order will, then, be
\begin{equation}
P_{conf}(eB/\sigma_0)=c_0^4 N_c^2\,\sigma_0^2\left[1-\tilde{\alpha}\left(\frac{eB}{\sigma_0}\right) \frac{N_{pairs}(N_f)}{N_c}\right]\,
\end{equation}
with $N_{pairs}(N_f)/N_c \ll 1$ being the number of pairs of quark flavors with electric charges $\left\{ (N_{c}-1)/N_{c},-1/N_{c}\right\}$ in units of the fundamental charge. Only the largest ($\sim N_{c}^{0}$) charge in each pair contributes to leading order in $N_{f}/N_{c}$.

In the deconfined phase, the $N_c^2$ contribution to the pressure is again $P_{glue}(T)=N_c^2 T^4 \,c_{SB}^4\,f_{glue}(T/\sqrt{\sigma_0})$ but the $N_f N_c$ flavor correction $P_{quark}$ feels directly the effects of the magnetic field. In fact, the regularized contribution \cite{comment2} of the massless quarks to the pressure is $P_{quark}(T,eB)=N_c \,N_{pairs}(N_f) \,T^4\,c_{q\,SB}^4\,\tilde{f}_{quark}(T/\sqrt{\sigma_0},eB/T^2)$.

Notice that the function $\tilde{f}_{quark}$ is positive definite and must increase monotonically with $T$ for a fixed value of $eB$ until it goes to 1 in the high temperature limit $T \gg \sqrt{\sigma_0}$, $eB$. Given our previous analysis for the case where $N_f\neq 0$ and $B=0$, one should expect that the critical temperature as a function of the magnetic field, $T_c(eB)$, must decrease with respect to pure glue value $T_c^{(0)}$ by an amount of $\mathcal{O}(N_f/N_c)$.

This can be seen directly by equating the pressures at $T_c$ 
\begin{eqnarray}
&& c_0^4 N_c^2\,\sigma_0^2\left[1-\tilde{\alpha}\left(\frac{eB}{\sigma_0}\right) \frac{N_{pairs}(N_f)}{N_c}\right] \nonumber \\ 
&=& N_c^2 T_c^4\,c_{SB}^4\, f_{glue}\left(\frac{T_c}{\sqrt{\sigma_0}}\right) \nonumber \\ &+& N_c \,N_{pairs}(N_f) \,T_c^4\,c_{q\,SB}^4\,\tilde{f}_{quark}\left(\frac{T_c}{\sqrt{\sigma_0}},\frac{eB}{T_c^2}\right)
 \label{equatingpressureswithB}
\end{eqnarray}
and noticing that, since the left-hand side of the equation above is fixed, the addition of the quark contribution on the right-hand side must lead to a decrease of the critical temperature by an amount of order $N_f/N_c$. In fact, the solution of the equation above for $T_c(eB)$, to leading order in $N_f/N_c$, is 
\begin{equation}
\frac{T_c(eB)}{\sqrt{\sigma_0}}\,f_{glue}^{1/4}\left(\frac{T_c(eB)}{\sqrt{\sigma_0}}\right)=\frac{c_2(N_{pairs},eB)}{c_{SB}}\,,
\label{newTcmag}
\end{equation}
where we defined
\begin{eqnarray}
&&c_2(N_{pairs},eB) \equiv c_0\left[1-\frac{1}{4}\tilde{\alpha}\left(\frac{eB}{\sigma_0}\right) \frac{N_{pairs}(N_f)}{N_c}\right]\nonumber \\ &\times& \left[1-\frac{1}{4}\frac{N_{pairs}(N_f)}{N_c}\frac{c_{q\,SB}^4\,\tilde{f}_{quark}\left(\frac{T_c^{(0)}}{\sqrt{\sigma_0}},\frac{eB}{T_c^{(0)\,2}}\right)}
{c_{SB}^4\,f_{glue}\left(\frac{T_c^{(0)}}{\sqrt{\sigma_0}}\right)}\right]\,.
\label{c2B}
\end{eqnarray}   

Since $c_2(N_{pairs},eB)< c_0$, the same arguments used before show that $T_c(eB)/T_c^{(0)}< 1$ by an amount $\sim N_f/N_c$. Therefore, one concludes that, in the presence of an external magnetic field, the deconfinement critical temperature decreases with respect to its value for pure glue in the large $N_c$ limit 
of QCD. Whether $T_c(eB)$ is also lower than the critical temperature in the presence of $N_{f}/N_{c}$ flavors of  massless quarks at $B=0$, $T_c^{(1)}$, requires that $c_2(N_{pairs},eB)< c_1$. This can be rewritten as a condition on the derivatives with respect to $B$ of the quark pressure, i.e. the magnetization $M(T_c,eB)$, and of the modification of the string tension, $\partial_B\tilde{\alpha}$:
$M(T_c,eB) > {\rm max} \left\{0\,,\,-c_{SB}^4f_{glue}\partial_B\tilde{\alpha}\right\}$.
This occurs if the flavor contribution behaves paramagnetically, with positive magnetization $M(T_c,eB)$ that is sufficiently large.

For a free gas implementation of the deconfined phase $f_{glue}=1$ and, in the limit of strong magnetic fields $eB/T^2  \gg 1$, one finds that $\tilde{f}_{quark}\sim eB/T_c^2$ \cite{Fraga:2012fs}. Assuming that the magnetic effects on the string tension are negligible, we may set $\tilde{\alpha}=\alpha$. Thus, in this case the magnetic suppression of the deconfinement critical temperature goes like $eB\,N_{pairs}/(N_c\,\sigma_0)$. In fact, this simple implementation in the limits of low and high magnetic fields provides a scenario in which the slope in $T_c(eB)$ decreases for large fields, as illustrated in Fig.\ 1.

\begin{figure}[h!]
\hspace{-1.5cm}
\includegraphics[width=7cm,angle=90]{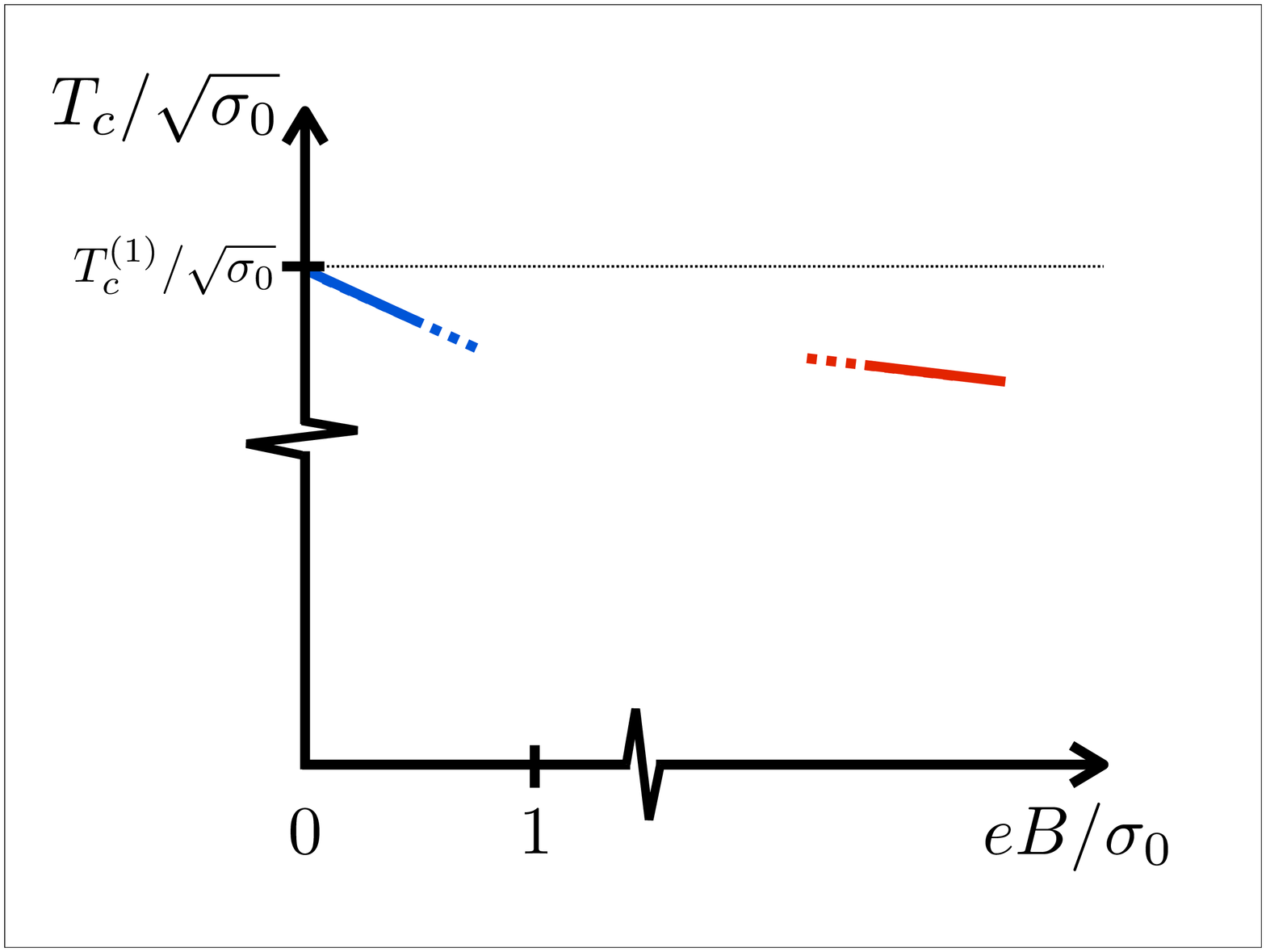}
\caption{Cartoon of the $T_c \times eB$ phase diagram in the large $N_{c}$ limit, using the approximation of 
free deconfined quarks and gluons and the assumption that magnetic effects on the string tension 
are negligible, i.e. $\tilde\alpha=\alpha$.
}
\end{figure}

An eventual saturation of $T_c$ as a function of $eB$, as observed on the lattice \cite{Bali:2011qj} and in model calculations \cite{Fraga:2012fs}, cannot be obtained using the limits discussed in this paper in a general fashion. As mentioned above, the implications of large $N_c$ estimates to the actual QCD phase diagram must be taken with great caution. The specific form of $T_c$ as a function of $eB$ depends on the non-perturbative functions $f_{glue}$, $\tilde{f}_{quark}$, and $\tilde{\alpha}$. In fact, in the large $N_c$ limit, our results indicate that $T_c(eB)$ can only be a flat curve if $\tilde{f}_{quark}$ and $\tilde{\alpha}$ are such that $M(T_c,eB)$ is positive but vanishes for large fields. In this scenario, a reasonable explanation for the nearly flat curve found in the $N_{c} = N_{f} = 3$ lattice study performed in Ref.\ \cite{Bali:2011qj} is a net cancellation effect that occurs for sufficiently large fields due to a magnetic field dependent contribution to the pressure below the phase transition (which in the physical case includes the dynamics of mesons).

{\it Quark mass effects.} 
When $m_q \neq 0$ the pressure of the confined phase is increased by
 the quark contribution to the vacuum trace anomaly, $N_c N_f m_q (-\langle\bar{q} q\rangle)$, where we used the fact that the quark condensate is negative. This is equivalent to a small positive shift of $c_0$ and, to leading order in $N_f/N_c$, the confined phase pressure when $eB=0$ is $P_{conf}=c_{m_q}^4 N_c^2 \sigma_0^2$, where
\begin{eqnarray}
c_{m_q}=c_0\left(
1-\frac{\alpha_{m_q}}{4} \frac{N_f}{N_c}+\frac{1}{4}\frac{N_f}{N_c}\frac{m_q}{\sqrt{\sigma_0}}\frac{(-\langle\bar{q} q\rangle)}{c_0^4\sigma_0^{3/2}}
\label{cmq}
\right)\,.
\end{eqnarray}
Here $\alpha_{m_q}$ (assumed to be positive) includes possible quark mass effects on the $\alpha$ coefficient. In the deconfined phase only the quark pressure will be affected by the quark mass effects, decreasing e.g. in perturbation theory  \cite{kapustagale}. In a large temperature expansion, we may write: $f_{quark}\mapsto f_{quark}-c_3 m_q^2/T^2$, where $c_3$ is positive. Therefore, the critical temperature computation in this massive case follows the same steps that led to Eqs. (\ref{c1Nf}) and (\ref{newTc}), with the substitution $f_{quark}\Big|_{m_q=0}\mapsto f_{quark}\Big|_{m_q=0}-c_3 m_q^2/T^2<f_{quark}\Big|_{m_q=0}$. As a consequence, $c_1(N_f,m_q)>c_1(N_f,m_q=0)$ and $T_c^{(m_q)}$ is higher than its massless counterpart, $T_c^{(1)}$. Interestingly enough, however, the corrections to $f_{quark}$ are
respectively $\sim (m_q/T_c^{(0)})^2$, being extremely small for reasonable values of quark masses, $m_q\ll \sqrt{\sigma_0}, T_c^{(0)}$. Therefore, in this large $N_c$ regime, we find that the critical temperature as a function of $m_q$ is essentially flat. Similar behavior has been observed on the lattice for $SU(3)$ \cite{Cheng:2007jq,Karsch:2000kv}.

Of course, the explicit dependence of $T_c$ with respect to 
the quark mass (or equivalently the pion mass) will also depend on the details of the functions $f_{glue}$, $f_{quark}$, $\alpha$ (which may acquire an extra dependence on the quark mass) as well as the quark condensate. In the study performed in \cite{Fraga:2008be,leticia-tese} within an effective model implementation of the $N_c=3$ and $N_f=2$ deconfined phase, $T_c/\sqrt{\sigma_0}$ was found to be nearly constant with respect to variations in the pion mass.

In the presence of a magnetic field, the quark condensate and its influence on $T_c$ are unaltered at this order in $N_f/N_c$, while the quark pressure receives magnetic contributions, becoming $\hat{f}_{quark}(T/\sqrt{\sigma_0},m_q/T,eB/T^2)$. Therefore, the critical temperature $T_c^{(2,m_q)}$ is the solution of Eq. (\ref{newTcmag}) with  $c_2$ replaced by
\begin{eqnarray}
&&\frac{c_2(N_{pairs},eB,m_q)}{c_{m_q}(N_{pairs},eB,m_q)} = 
\left[1 - \frac{1}{4} \right. \nonumber \\ &\times& \left.\frac{N_{pairs}(N_f)}{N_c} \frac{c_{q\,SB}^4\,\hat{f}_{quark}\left(\frac{T_c^{(0)}}{\sqrt{\sigma_0}},\frac{m_q}{T_c^{(0)}},\frac{eB}{T_c^{(0)\,2}}\right)}
{c_{SB}^4\,f_{glue}\left(\frac{T_c^{(0)}}{\sqrt{\sigma_0}}\right)}\right]\,
\end{eqnarray}
where $c_{m_q}(N_{pairs},eB,m_q)$ is the corresponding generalization of $c_{m_q}$ in Eq.\ (\ref{cmq}) that takes into account magnetic field effects. In this more complicated scenario there will be a competition between mass and magnetic effects and it is hard to obtain even a qualitative estimate of the general behavior of the critical temperature as a function of $eB$. If, however, the term that is most sensitive to the magnetic field is $\hat{f}_{quark}$, then if this term is paramagnetic then the critical temperature would assume values that are lower than $T_c^{(m_q)}$ as one varies the magnetic field. 

{\it Final comments.} It would be interesting to extend the discussion about the magnetic effects on the deconfinement critical temperature to the Veneziano limit of QCD. In this case, one could also study whether chiral symmetry restoration coincides with the deconfinement transition when $N_f$, $N_c \to \infty$ in the presence of an external magnetic field.        

{\it Acknowledgments:} 
E.S.F. and J.N. acknowledge the hospitality of Departamento de F\'\i sica Te\'orica at UERJ, 
where most of this work has been conducted. The authors thank 
M. Chernodub, M. D'Elia, R. D. Pisarski, and I. Shovkovy  for comments and discussions.
This work was partially supported by CAPES, CNPq, FAPERJ, FAPESP and FUJB/UFRJ.



\begin{thebibliography}{9}

\bibitem{magnetic-HIC}
 V.~Skokov, A.~Y.~Illarionov and V.~Toneev,
 Int.\ J.\ Mod.\ Phys.\  A {\bf 24}, 5925 (2009);
%
 V.~Voronyuk, V.~D.~Toneev, W.~Cassing, E.~L.~Bratkovskaya, V.~P.~Konchakovski and S.~A.~Voloshin,
 Phys.\ Rev.\ C {\bf 83}, 054911 (2011); 
%
  A.~Bzdak and V.~Skokov,
  Phys.\ Lett.\ B {\bf 710}, 171 (2012); 
  W.~-T.~Deng and X.~-G.~Huang,
  Phys.\ Rev.\ C {\bf 85}, 044907 (2012).

\bibitem{Agasian:2008tb}
 N.~O.~Agasian and S.~M.~Fedorov,
 Phys.\ Lett.\  B {\bf 663}, 445 (2008).

\bibitem{Fraga:2008qn}
 E.~S.~Fraga and A.~J.~Mizher,
 Phys.\ Rev.\  D {\bf 78}, 025016 (2008);
%
 A.~J.~Mizher, M.~N.~Chernodub and E.~S.~Fraga,
 Phys.\ Rev.\ D {\bf 82}, 105016 (2010).
 
\bibitem{Menezes:2008qt} 
S.S. Avancini, D.P. Menezes, M.B.Pinto and C. Provid\^{e}ncia, Phys. 
Rev. D {\bf 85}, 091901 (2012);
  G.~N.~Ferrari, A.~F.~Garcia and M.~B.~Pinto,
  arXiv:1207.3714 [hep-ph].

\bibitem{Boomsma:2009yk}
J.~K.~Boomsma and D.~Boer,
 Phys.\ Rev.\  D {\bf 81}, 074005 (2010).

\bibitem{Fukushima:2010fe}
 K.~Fukushima, M.~Ruggieri and R.~Gatto,
 Phys.\ Rev.\  D {\bf 81}, 114031 (2010);
%
 R.~Gatto and M.~Ruggieri,
 Phys.\ Rev.\ D {\bf 82}, 054027 (2010);
%
 R.~Gatto and M.~Ruggieri,
 Phys.\ Rev.\ D {\bf 83}, 034016 (2011).
 
\bibitem{Avancini:2011zz} 
  S.~S.~Avancini, D.~P.~Menezes and C.~Providencia,
  Phys.\ Rev.\ C {\bf 83}, 065805 (2011).
  
\bibitem{Kashiwa:2011js} 
  K.~Kashiwa,
  Phys.\ Rev.\ D {\bf 83}, 117901 (2011).
  
\bibitem{Chatterjee:2011ry} 
  B.~Chatterjee, H.~Mishra and A.~Mishra,
  Phys.\ Rev.\ D {\bf 84}, 014016 (2011).
  
\bibitem{Andersen:2011ip} 
  J.~O.~Andersen and R.~Khan,
  Phys.\ Rev.\ D {\bf 85}, 065026 (2012);
  %
  J.~O.~Andersen and A.~Tranberg,
  arXiv:1204.3360 [hep-ph].
 
\bibitem{Skokov:2011ib} 
  V.~Skokov,
  Phys.\ Rev.\ D {\bf 85}, 034026 (2012).
  
\bibitem{Fukushima:2012xw} 
  K.~Fukushima and J.~M.~Pawlowski,
  arXiv:1203.4330 [hep-ph].

\bibitem{Chernodub:2010qx}
 M.~N.~Chernodub,
 Phys.\ Rev.\ D {\bf 82}, 085011 (2010);
%
 M.~N.~Chernodub,
 Phys.\ Rev.\ Lett.\  {\bf 106}, 142003 (2011);
%
  M.~N.~Chernodub,
  Int.\ J.\ Mod.\ Phys.\ A {\bf 27}, 1260003 (2012).
  
\bibitem{lattice-delia}
M.~D'Elia, S.~Mukherjee and F.~Sanfilippo,
 Phys.\ Rev.\  D {\bf 82}, 051501 (2010);
%
 M.~D'Elia and F.~Negro,
 Phys.\ Rev.\ D {\bf 83}, 114028 (2011).

\bibitem{Bali:2011qj} 
  G.~S.~Bali, F.~Bruckmann, G.~Endrodi, Z.~Fodor, S.~D.~Katz, S.~Krieg, A.~Schafer and K.~K.~Szabo,
  JHEP {\bf 1202}, 044 (2012).
 
\bibitem{Gusynin:1994xp}
  V.~P.~Gusynin, V.~A.~Miransky and I.~A.~Shovkovy,
  Phys.\ Lett.\  B {\bf 349}, 477 (1995);
%
  V.~P.~Gusynin, V.~A.~Miransky and I.~A.~Shovkovy,
  Nucl.\ Phys.\  B {\bf 462}, 249 (1996);
%
%
  G.~W.~Semenoff, I.~A.~Shovkovy and L.~C.~R.~Wijewardhana,
  Phys.\ Rev.\  D {\bf 60}, 105024 (1999); 
%
  V.~A.~Miransky and I.~A.~Shovkovy,
  Phys.\ Rev.\   D {\bf 66}, 045006 (2002);
    I.~A.~Shovkovy,
arXiv:1207.5081 [hep-ph].
  
\bibitem{Bali:2012zg} 
  G.~S.~Bali, F.~Bruckmann, G.~Endrodi, Z.~Fodor, S.~D.~Katz and A.~Schafer,
  arXiv:1206.4205 [hep-lat].
 
\bibitem{Fraga:2012fs} 
  E.~S.~Fraga and L.~F.~Palhares,
  Phys.\ Rev.\  D {\bf 86}, 016008 (2012).
  
\bibitem{Fraga:2008be} 
  E.~S.~Fraga, L.~F.~Palhares and C.~Villavicencio,
  Phys.\ Rev.\ D {\bf 79}, 014021 (2009);
%
  L.~F.~Palhares, E.~S.~Fraga and C.~Villavicencio,
  Nucl.\ Phys.\ A {\bf 820}, 287C (2009).
  
 \bibitem{leticia-tese}
 L.~F.~Palhares, {\it Exploring the Different Phase Diagrams of Strong Interactions} 
 (PhD Thesis, Federal University of Rio de Janeiro, 2012),  arXiv:1208.0574 [hep-ph].
 
\bibitem{'tHooft:1973jz} 
  G.~'t Hooft,
  Nucl.\ Phys.\ B {\bf 72}, 461 (1974).
  
\bibitem{'tHooft:1974hx} 
  G.~'t Hooft,
  Nucl.\ Phys.\ B {\bf 75}, 461 (1974).
  
\bibitem{Witten:1979kh} 
  E.~Witten,
  Nucl.\ Phys.\ B {\bf 160}, 57 (1979).
  
\bibitem{Manohar:1998xv} 
  A.~V.~Manohar,
  hep-ph/9802419.
  
\bibitem{Teper:2008yi} 
  M.~Teper,
  PoS LATTICE {\bf 2008}, 022 (2008).
  
\bibitem{Boyd:1996bx} 
  G.~Boyd, J.~Engels, F.~Karsch, E.~Laermann, C.~Legeland, M.~Lutgemeier and B.~Petersson,
  Nucl.\ Phys.\ B {\bf 469}, 419 (1996).
  
\bibitem{muitaslattice}
  B.~Lucini, M.~Teper, and U.~Wenger, Phys.\ Lett.\ B {\bf 545},
197 (2002); J.\ High Energy Phys.\ 01 (2004) 061; K.~Holland, M.~Pepe, and
U.-J.~Wiese, Nucl.\ Phys.\ B {\bf 694}, 35 (2004); Nucl.\ Phys.\ B,
Proc.\ Suppl.\ {\bf 129�130}, 712 (2004); M.~Pepe, Nucl.\ Phys.\
B, Proc.\ Suppl.\ {\bf 141}, 238 (2005).

\bibitem{Lucini:2005vg} 
  B.~Lucini, M.~Teper and U.~Wenger,
  JHEP {\bf 0502}, 033 (2005).

\bibitem{Borsanyi:2012ve} 
  S.~.Borsanyi, G.~Endrodi, Z.~Fodor, S.~D.~Katz and K.~K.~Szabo,
  JHEP {\bf 1207}, 056 (2012).
  
\bibitem{Lucini:2012wq} 
  B.~Lucini, A.~Rago and E.~Rinaldi,
  Phys.\ Lett.\ B {\bf 712}, 279 (2012).

\bibitem{morelattice}
B.~Bringoltz and M.~Teper, Phys.\ Lett.\ B {\bf 628}, 113 (2005);
S.~Datta and S.~Gupta, Nucl.\ Phys.\ A {\bf 830}, 749c (2009).

\bibitem{Panero:2009tv} 
  M.~Panero,
  Phys.\ Rev.\ Lett.\  {\bf 103}, 232001 (2009).

\bibitem{kapustagale}
  J.~Kapusta and C.~Gale, 
  {\it Finite Temperature Field Theory: Principles and Applications} 
  (Cambridge University Press, 2006).

\bibitem{comment1} This can be proven as follows. Consider the equation $x^4 g(x)=c$ for $c>0$, $x > 0$, with $g(x)$ being a positive definite, monotonically increasing
function of $x$ that is limited when $x \in [0,\infty)$. Let $x_c \equiv x(c) > 0$ be a nontrivial solution of this equation. $x_c$ vanishes only if $c$ also does. Then,
taking the derivative of this equation with respect to $c$ one obtains that $x'(c)=\left[4c/x_c+c \,g'(x_c)/g(x_c)\right]^{-1} \geq 0$, $\forall c>0$. Therefore, the 
nontrivial solution $x_c$ must increase with $c$. 
  
  \bibitem{Karsch}
  � F.~Karsch, E.~Laermann and A.~Peikert,
� 
� Nucl.\ Phys.\ B {\bf 605}, 579 (2001).
  
\bibitem{Aoki:2006we} 
  Y.~Aoki, G.~Endrodi, Z.~Fodor, S.~D.~Katz and K.~K.~Szabo,
  Nature {\bf 443}, 675 (2006).
  
\bibitem{Aoki:2006br} 
  Y.~Aoki, Z.~Fodor, S.~D.~Katz and K.~K.~Szabo,
  Phys.\ Lett.\ B {\bf 643}, 46 (2006).
  
\bibitem{Bazavov:2011nk} 
  A.~Bazavov {\it et al.} [HotQCD Collaboration],
  Phys.\ Rev.\ D {\bf 85}, 054503 (2012).
  
\bibitem{Agasian:1999sx} 
  N.~O.~Agasian and I.~A.~Shushpanov,
  Phys.\ Lett.\ B {\bf 472}, 143 (2000).
  
\bibitem{Agasian:2000hw} 
  N.~O.~Agasian,
  Phys.\ Lett.\ B {\bf 488}, 39 (2000).
  
\bibitem{Cohen:2007bt} 
  T.~D.~Cohen, D.~A.~McGady and E.~S.~Werbos,
  Phys.\ Rev.\ C {\bf 76}, 055201 (2007).
  
\bibitem{comment2} The contribution coming from the magnetic field that does not depend on either $m_q$ (which is set to zero) and $T$ \cite{Fraga:2012fs} must be subtracted out in this analysis. This corresponds to a choice of the renormalization scale in the electromagnetic sector and presents important physical consequences, as discussed in detail in Ref. \cite{Fraga:2012fs}. 

\bibitem{Cheng:2007jq}
 M.~Cheng {\it et al.},
 Phys.\ Rev.\  D {\bf 77}, 014511 (2008).

\bibitem{Karsch:2000kv}
 F.~Karsch, E.~Laermann and A.~Peikert,
 Nucl.\ Phys.\  B {\bf 605}, 579 (2001); 
 F.~Karsch, K.~Redlich and A.~Tawfik,
 Eur.\ Phys.\ J.\  C {\bf 29}, 549 (2003).

\bibitem{Veneziano:1976wm} 
  G.~Veneziano,
  Nucl.\ Phys.\ B {\bf 117}, 519 (1976).
 

\end{thebibliography}
\end{document}